\documentclass{iucrjournals-arxiv}

\usepackage{amsmath,amssymb}
\usepackage{listings}
\definecolor{coderule}{RGB}{200,205,210}
\definecolor{codekw}{RGB}{0,0,180}
\definecolor{codecomment}{RGB}{60,120,60}
\definecolor{codestr}{RGB}{160,80,0}
\newfloat{listing}{tbp}{lol} 
\floatname{listing}{Listing}

\usepackage{siunitx}
\usetikzlibrary{arrows.meta,calc,positioning}

\hypersetup{colorlinks=true,linkcolor=black,citecolor=blue!50!black,urlcolor=blue}
\usepackage[capitalise,noabbrev]{cleveref}

\newcommand{\vh}{\mathbf{h}}
\newcommand{\vx}{\mathbf{x}}
\newcommand{\vH}{\mathbf{H}}
\newcommand{\vR}{\mathbf{R}}
\newcommand{\vk}{\mathbf{k}}
\newcommand{\vq}{\mathbf{q}}
\newcommand{\vu}{\mathbf{u}}
\newcommand{\vC}{\mathbf{C}}
\newcommand{\vU}{\mathbf{U}}
\newcommand{\ve}{\mathbf{e}}
\newcommand{\kB}{k_{\mathrm{B}}}
\newcommand{\T}{^{\mathsf{T}}}

\newcommand{\ddPDF}{3D\(\Delta\)PDF}

\providecommand{\figw}{0.5\linewidth}

\title{X-ray Thermal diffuse scattering from real-space displacement
correlations}
\author[a]{Benjamin Fahl}\author[a]{Jonathan Bulled}\author[b]{Artem Korshunov}\author[c]{Dmitry Chernyshov}\author[d]{Yevheniia Kholina}\author[a,e]{Arkadiy Simonov\IUCrCemaillink{arkadiy.simonov@mat.ethz.ch}}

\affil[a]{Department of Materials, ETH Zurich, Zurich, Switzerland}
\affil[b]{ESRF -- The European Synchrotron, ID28, Grenoble, France}
\affil[c]{Swiss--Norwegian Beamlines (SNBL), ESRF, Grenoble, France}
\affil[d]{Department of Chemistry, University of Oxford, Oxford, United Kingdom}
\affil[e]{Department of Chemistry, Aarhus University, Aarhus, Denmark}

\begin{document}
\maketitle

\begin{synopsis}
We propose a fast method for calculating X-ray thermal diffuse scattering 
based on the parameterization in three-dimensional pair distribution function 
space.
\end{synopsis}

\begin{abstract}
  We introduce a method for calculating X-ray thermal diffuse
  scattering based on the three-dimensional difference pair
  distribution function (3D-$\Delta$PDF). Within the harmonic
  approximation the method is exact, as it includes all orders of
  multi-phonon scattering. A single Fourier transform delivers the
  diffuse intensity over large volumes of reciprocal space. We tested
  the method against experimental diffuse scattering measured on a
  silicon single crystal. With nothing refined beyond a scale and a
  background, the calculation reproduces the measured intensity to an
  $R_2$ residual below 5\%. Because the method uses the same
  pair-correlation language as the established \textsc{Yell} program,
  thermal and static correlated disorder enter the analysis on equal
  footing.
\end{abstract} 

\begin{figure*}[!t]
  \centering
  \includegraphics[width=\textwidth]{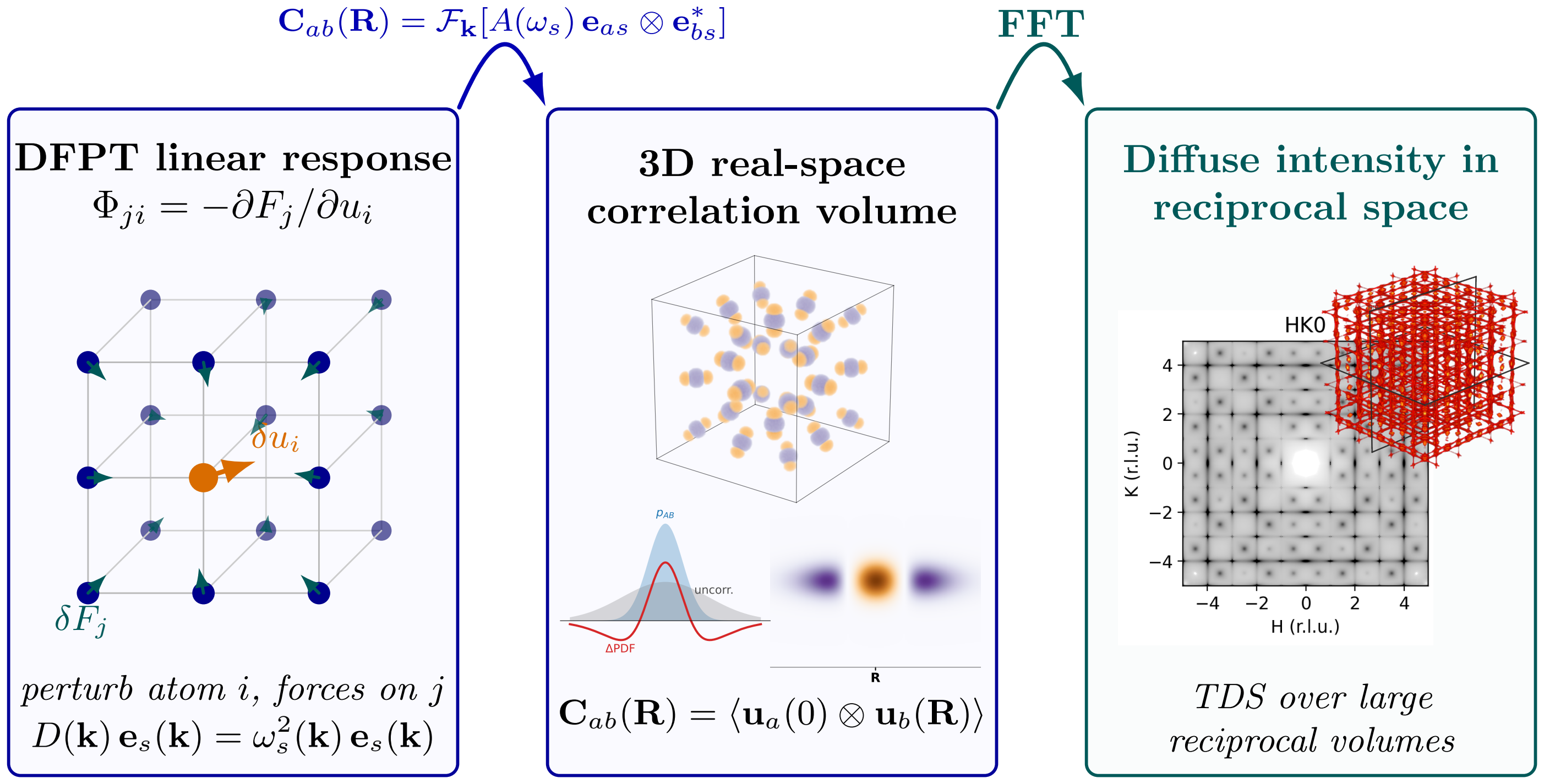}
  \caption{Real-space correlation approach: pair displacement correlations form a
    3D real-space volume that is FFT'd to obtain the diffuse intensity.\label{fig:tds_method_schematic}}
\end{figure*}
 \section{Introduction}
\enlargethispage{\baselineskip}

Thermal diffuse scattering (TDS) is a continuous signal underneath the Bragg
peaks in single-crystal diffraction. It is an energy-integrated sum over every
phonon mode of the crystal, weighted by thermal occupation. In the 1940s it was
the primary probe of lattice dynamics, and Laval, Born and James
\cite{Laval1939,born1942theoretical,James1948} used it to measure phonon
dispersions and elastic constants of various inorganic materials. Shortly after,
TDS fell out of fashion for several decades, eclipsed by energy-resolved
inelastic neutron and later inelastic X-ray scattering (INS and IXS). Recent
years have seen a modest revival, because the instrumentation for TDS is simpler than
for inelastic measurements and the full reciprocal-space volume can be collected quickly. The 
method extends naturally to high-pressure
studies~\cite{xuDeterminationPhononDispersion2005,wehingerFullElasticity2017,bosakModelfreeReconstructionLattice2008,grevePronouncedNegative2010,liStructuralRelationship2011},
and has recently been used to investigate the lattice dynamics of thermoelectric materials
\cite{sangiorgio2018correlated, holm2021anharmonicity}, halide perovskites 
\cite{dubajic2025dynamic}, and even macromolecular crystals~\cite{wall2014conformational, meisburger2020diffuse}.
Time-resolved experiments extend the method further, for instance to non-equilibrium lattice 
dynamics such as phonon emission after photoexcitation~\cite{pan2025momentum}.

While measuring TDS has become routine, calculating it accurately and efficiently remains 
challenging, especially over large volumes of reciprocal space.
The established methods fall into two branches. The first
is the Laval-Born-James approach, which truncates the phonon expansion at one- and
two-phonon scattering~\cite{xuXrayThermalDiffuse2010, mirone_ab2tds}.
These calculations are fast and easy to interpret, and they recover elastic constants~\cite{wehingerFullElasticity2017} and
in simple cases phonon dispersions~\cite{xuXrayThermalDiffuse2010}. The two-phonon approximation
is not always enough, though. At large scattering vectors and at modestly high
temperatures the higher multi-phonon terms become important and the truncation error starts to
show. Going beyond two phonons following this approach is impractical. The one-phonon term is 
cheap as it is evaluated point by point. The two-phonon term is more expensive as it requires 
an integral over the first Brillouin zone, which makes the calculation four to five orders of 
magnitude slower than the one-phonon calculation. Each further order adds another such integral,
so the cost increases again by the same factor. To our knowledge, no Laval-Born-James calculation has gone beyond two phonons.

The second branch avoids the expansion altogether. Instead of summing phonon orders, 
one can build a supercell in which the atoms are displaced according to the thermal 
distribution, and then calculate the diffuse scattering from that supercell. Such a calculation 
naturally includes all phonon orders, and has been widely used to compute the diffuse
scattering from 
macromolecular dynamical trajectories~\cite{wall2014conformational, meisburger2020diffuse}. The drawback of such an approach is the sampling noise.
Molecular dynamics samples the thermal distribution at random, so many snapshots
must be averaged before the diffuse scattering converges. The special displacement method of 
Zacharias and Giustino~\cite{zacharias2021multiphonon} is a clever way to reduce this noise. 
Instead of averaging over random snapshots, they build a single supercell whose displacements 
are chosen carefully to reproduce the thermal average. This gives a good approximation
even from a single supercell, but the result still carries a statistical error.

In this paper we introduce a third route, based on the difference pair distribution
function ($\Delta$PDF) method \cite{weberThreedimensionalPairDistribution2012}. From a
first-principles phonon calculation we obtain the covariances of the atomic
displacements, and from these we build the 3D-$\Delta$PDF map. A single Fourier transform
of that volume gives the diffuse intensity in the infinite-phonon approximation, with no
truncation and no sampling error. The method has a further advantage: it treats thermal
and static disorder on equal
footing, because both are expressed in the same pair-correlation language that the
software \textsc{Yell} uses for static correlated
disorder~\cite{simonovYellComputerProgram2014}, so a sample showing both can be
treated at once. Thanks to a fast Fourier transform algorithm, the method stays efficient over
large volumes of reciprocal space.

The rest of the paper is organized as follows. Section~\ref{sec:background} sets out the
theory, Section~\ref{sec:implementation} describes the implementation, and
Section~\ref{sec:validation} compares the calculated diffuse scattering with experimental
data for silicon.
 \section{Background}\label{sec:background}

To keep the paper self-contained, we review the standard kinematic theory of thermal diffuse 
scattering here, mainly to fix our notation. The treatment follows the classical
accounts~\cite{born1942theoretical,
warrenXrayDiffraction1969,James1948,willisPryorThermalVibrations1975},
with notation adapted to the
3D-$\Delta$PDF
framework~\cite{weberThreedimensionalPairDistribution2012,simonovYellComputerProgram2014}
used in \textsc{Yell}.

Throughout we use the crystallographic convention: the scattering
vector $\vh$ is measured in reciprocal-lattice units, real-space
positions and displacements are expressed in the crystal (fractional)
basis, and the displacement covariances $\vU_\kappa$ and
$\vC_{\kappa\kappa'}$ are referred to that same basis, so that the
products $\vh\!\cdot\!\mathbf{r}$ and
$\vh\T\vC\,\vh$ are dimensionless and
Bragg reflections fall at integer $\vh$. Fourier transforms use
the crystallographic kernel $e^{2\pi i\,\vh\cdot\mathbf{r}}$.

\subsection{Scattering from correlated thermal displacements}

At finite temperature each atom $\kappa$ in cell $n$ vibrates about its equilibrium site, 
so its instantaneous position is
\begin{equation}
  \mathbf{r}_{n \kappa} = \vR_n + \vx_\kappa + \vu_{n \kappa},
\end{equation}
where $\vR_n$ is the lattice vector of cell $n$, with integer
coordinates in the crystal basis; $\vx_\kappa$ is the average basis
position; and $\vu_{n \kappa}$ is the thermal displacement.

The instantaneous scattering amplitude at scattering vector $\vh$ is
\begin{equation}
  A(\vh) =
  \sum_{n,\kappa}
  f_\kappa(\vh)
  \exp\!\bigl[
    2\pi i\, \vh\!\cdot\!
    (\vR_n + \vx_\kappa + \vu_{n \kappa})
  \bigr],
\end{equation}
where $f_\kappa(\vh)$ is the atomic form factor.
The thermally averaged intensity is
\begin{align}
  I(\vh)
  &=
  \bigl\langle A(\vh) A^{*}(\vh) \bigr\rangle
  \nonumber\\
  &=
  \sum_{n,m}\sum_{\kappa,\kappa'}
  f_\kappa(\vh) f_{\kappa'}^{*}(\vh)
  \nonumber\\
  &\quad\times
  \exp\!\bigl[
    2\pi i\, \vh\!\cdot\!
    (\vR_n - \vR_m + \vx_\kappa - \vx_{\kappa'})
  \bigr]
  \nonumber\\
  &\quad\times
  \bigl\langle
  \exp\!\bigl[
    2\pi i\, \vh\!\cdot\!(\vu_{n \kappa} - \vu_{m \kappa'})
  \bigr]
  \bigr\rangle.
  \label{eq:thermally_averaged_intensity}
\end{align}

\noindent
Because the displacement field is Gaussian under the harmonic
approximation, the thermal average of the phase factor in
Eq.~\eqref{eq:thermally_averaged_intensity} reduces to an exponential of its
variance~\cite{born1942theoretical}:
\begin{equation}
  \bigl\langle e^{2\pi i\, \vh\cdot(\vu_{n \kappa}-\vu_{m \kappa'})}
  \bigr\rangle
  =
  \exp\!\left[
    -2\pi^2
    \bigl\langle
    (\vh\!\cdot\!(\vu_{n \kappa}-\vu_{m \kappa'}))^2
    \bigr\rangle
  \right].
\end{equation}

\noindent
We define the displacement-covariance tensor
\begin{equation}
  \vC_{\kappa \kappa'}(\vR)
  =
  \bigl\langle
  \vu_{0 \kappa}\vu_{R \kappa'}\T
  \bigr\rangle,
  \qquad
  \vR=\vR_m-\vR_n,
\end{equation}
and the mean-square displacement tensor
\begin{equation}
  \vU_\kappa = \vC_{\kappa \kappa}(\mathbf{0}).
\end{equation}

\noindent
Using translational invariance, we find
\begin{multline}
  \bigl\langle
  (\vu_{n \kappa}-\vu_{m \kappa'})
  (\vu_{n \kappa}-\vu_{m \kappa'})\T
  \bigr\rangle
  \\
  =
  \vU_\kappa + \vU_{\kappa'}
  -\vC_{\kappa \kappa'}(\vR)
  -\vC_{\kappa' \kappa}(-\vR).
\end{multline}
Since
$\vh\T\vC_{\kappa' \kappa}(-\vR)\vh
=\vh\T\vC_{\kappa \kappa'}(\vR)\vh$,
the thermal average becomes
\begin{multline}
  \bigl\langle
  e^{2\pi i\, \vh\cdot(\vu_{n \kappa}-\vu_{m \kappa'})}
  \bigr\rangle
  \\
  =
  \exp\!\Bigl[
    -2\pi^2\,\vh\T
    (\vU_\kappa+\vU_{\kappa'})\vh
    \\
    +
    4\pi^2\,\vh\T\vC_{\kappa \kappa'}(\vR)\vh
  \Bigr].
\end{multline}

\noindent
Therefore, the full intensity can be written as
\begin{align}
  I(\vh)
  &=
  \sum_{n,m}\sum_{\kappa,\kappa'}
  f_\kappa(\vh) f_{\kappa'}^{*}(\vh)
  \nonumber\\
  &\quad\times
  \exp\!\bigl[
    2\pi i\, \vh\!\cdot\!
    (\vR_n-\vR_m+\vx_\kappa-\vx_{\kappa'})
  \bigr]
  \nonumber\\
  &\quad\times
  \exp\!\Bigl[
    -2\pi^2\,\vh\T(\vU_\kappa+\vU_{\kappa'})\vh
  \Bigr]
  \nonumber\\
  &\quad\times
  \exp\!\Bigl[
    4\pi^2\,\vh\T\vC_{\kappa \kappa'}(\vR_m-\vR_n)\vh
  \Bigr].
\end{align}

\subsection{Lattice-sum form}

Translational invariance lets us collapse the double cell sum to
a single sum over lattice vectors $\vR=\vR_m-\vR_n$
multiplied by the number of cells $N$:
\begin{align}
  I(\vh)
  &=
  N\sum_{\vR}\sum_{\kappa,\kappa'}
  f_\kappa(\vh) f_{\kappa'}^{*}(\vh)
  \nonumber\\
  &\quad\times
  \exp\!\bigl[
    2\pi i\, \vh\!\cdot\!
    (\vR+\vx_\kappa-\vx_{\kappa'})
  \bigr]
  \nonumber\\
  &\quad\times
  \exp\!\Bigl[
    -2\pi^2\,\vh\T(\vU_\kappa+\vU_{\kappa'})\vh
  \Bigr]
  \nonumber\\
  &\quad\times
  \exp\!\Bigl[
    4\pi^2\,\vh\T\vC_{\kappa \kappa'}(\vR)\vh
  \Bigr].
  \label{eq:full_intensity_realspace}
\end{align}

Equation~\eqref{eq:full_intensity_realspace} is exact within the harmonic-Gaussian 
description. It contains \emph{all} phonon orders through the exponential of the 
real-space displacement correlations. The next subsection
unpacks this exponential, separating Bragg from diffuse
scattering. The linearised one-phonon limit is given in the
Supplementary Information.

\subsection{Bragg-diffuse separation}

The Bragg and diffuse parts of the intensity correspond,
respectively, to the long-range correlated motion captured by
the constant term of the displacement-correlation exponential
and to its $\vR$-dependent remainder. This split is made
explicit by writing
\begin{multline}
  \exp\!\Bigl[
    4\pi^2\,\vh\T\vC_{\kappa \kappa'}(\vR)\vh
  \Bigr]
  \\
  =
  1 +
  \Bigl\{
    \exp\!\Bigl[
      4\pi^2\,\vh\T\vC_{\kappa \kappa'}(\vR)\vh
    \Bigr]
    -1
  \Bigr\},
\end{multline}
so that the total intensity separates as
$I(\vh) = I_{\mathrm{Bragg}}(\vh) +
I_{\mathrm{diff}}(\vh)$, with the constant term
generating the Bragg peaks and the bracketed remainder, which
vanishes as $|\vR|\!\to\!\infty$ for finite-range
correlations, generating the diffuse intensity.

The Bragg contribution is
\begin{align}
  I_{\mathrm{Bragg}}(\vh)
  &=
  N\sum_{\vR}\sum_{\kappa,\kappa'}
  f_\kappa(\vh) f_{\kappa'}^{*}(\vh)
  \nonumber\\
  &\quad\times
  \exp\!\bigl[
    2\pi i\, \vh\!\cdot\!
    (\vR+\vx_\kappa-\vx_{\kappa'})
  \bigr]
  \nonumber\\
  &\quad\times
  \exp\!\Bigl[
    -2\pi^2\,\vh\T(\vU_\kappa+\vU_{\kappa'})\vh
  \Bigr],
\end{align}
or equivalently at integer reciprocal-lattice vectors $\vH$,
\begin{equation}
  I_{\mathrm{Bragg}}(\vh)
  =
  N
  \sum_{\vH}
  \delta(\vh-\vH)
  \bigl|F_{\mathrm{avg}}(\vH)\bigr|^2,
\end{equation}
with average-structure factor
\begin{align}
  F_{\mathrm{avg}}(\vH)
  &=
  \sum_\kappa
  f_\kappa(\vH)
  \exp\!\bigl(2\pi i\, \vH\!\cdot\!\vx_\kappa\bigr)
  \nonumber\\
  &\quad\times
  \exp\!\Bigl(
    -2\pi^2\,\vH\T\vU_\kappa\vH
  \Bigr).
\end{align}

\noindent
The diffuse contribution is
\begin{align}
  I_{\mathrm{diff}}(\vh)
  &=
  N\sum_{\vR}\sum_{\kappa,\kappa'}
  f_\kappa(\vh) f_{\kappa'}^{*}(\vh)
  \nonumber\\
  &\quad\times
  \exp\!\bigl[
    2\pi i\, \vh\!\cdot\!
    (\vR+\vx_\kappa-\vx_{\kappa'})
  \bigr]
  \nonumber\\
  &\quad\times
  \exp\!\Bigl[
    -2\pi^2\,\vh\T(\vU_\kappa+\vU_{\kappa'})\vh
  \Bigr]
  \nonumber\\
  &\quad\times
  \Bigl[
    \exp\!\Bigl(
      4\pi^2\,\vh\T\vC_{\kappa \kappa'}(\vR)\vh
    \Bigr)-1
  \Bigr].
  \label{eq:all_order_diffuse}
\end{align}

The Bragg-diffuse separation introduced no approximation, so Eq.~\eqref{eq:all_order_diffuse} 
retains all phonon orders of Eq.~\eqref{eq:full_intensity_realspace}, now written for the 
diffuse intensity alone in terms of real-space displacement correlations.

\subsection{Connection to phonons}

The real-space covariance tensor can be obtained from phonons~\cite{xuXrayThermalDiffuse2010}
as
\begin{align}
  \vC_{\kappa \kappa'}(\vR,T)
  &= \frac{\hbar}{2 N_{\vq} \sqrt{M_\kappa M_{\kappa'}}}
  \sum_{\vq,\nu}
  \frac{
    \ve^{*}_{\kappa \nu}(\vq) \otimes
    \ve_{\kappa' \nu}(\vq)
  }{
    \omega_\nu(\vq)
  }
  \nonumber\\
  &\quad\times
  \coth\!\left(
    \frac{\hbar\omega_\nu(\vq)}{2 \kB T}
  \right)
  e^{2\pi i\, \vq\cdot\vR},
  \label{eq:covariance_from_phonons}
\end{align}
where $M_\kappa$ is the atomic mass, $\omega_\nu(\vq)$ is the
phonon frequency, and $\ve_{\kappa \nu}(\vq)$ is the
phonon eigenvector component for atom $\kappa$ in branch $\nu$, with the
resulting tensor referred to the crystal basis.
The on-site mean-square displacement tensor follows as
$\vU_\kappa(T) = \vC_{\kappa \kappa}(\mathbf{0},T)$.
Substituting Eq.~\eqref{eq:covariance_from_phonons} into
Eq.~\eqref{eq:all_order_diffuse} yields an all-order
phonon-based diffuse scattering formalism expressed entirely
through real-space pair correlations.

\begin{figure}[tb]
  \centering
  \includegraphics[width=\figw]{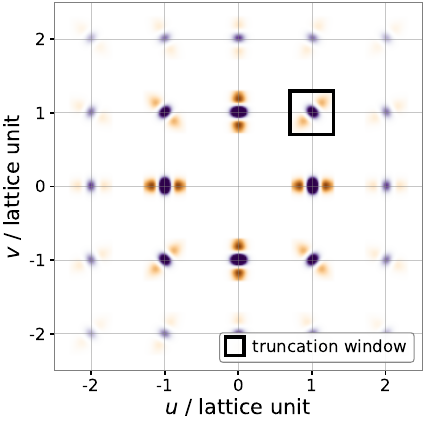}
  \caption{Real-space truncation in the 3D-$\Delta$PDF
    representation: a 2D slice of the pair displacement correlation
    map with one truncation window drawn (black box). The
    truncation size is an empirical parameter that trades accuracy
    for speed. Truncation is exaggerated for clarity; in practice
    windows of neighbouring signals
  overlap.\label{fig:pdf_truncation}}
\end{figure}

\subsection{Autocorrelation form}

The same diffuse intensity can be written as the Fourier
transform of the connected density autocorrelation,
\begin{equation}
  I_{\mathrm{diff}}(\vh)
  =
  \mathcal{F}
  \bigl[
    \langle\rho \star \rho\rangle
    -
    \langle\rho\rangle \star \langle\rho\rangle
  \bigr],
\end{equation}
where $\star$ denotes correlation and the term in brackets is the
three-dimensional difference pair distribution function
(3D-$\Delta$PDF), the real-space counterpart of the diffuse intensity.
In $\Delta$PDF space the diffuse signal takes a simple form of a sum of 
Gaussians, one per atom pair. Rather than evaluating the autocorrelation, 
\textsc{Yell} builds this sum directly and Fourier transforms it 
(\cref{sec:implementation}), the same construction it uses for static 
correlated disorder. Each Gaussian sits on an interatomic vector and is 
appreciable only near it (\cref{fig:pdf_truncation}), so it can be built 
inside a small box around that vector and skipped over the rest of the volume,
which is what makes the evaluation fast. The same speed-up has long been
exploited in structure-factor
evaluation~\cite{teneyckEfficientStructurefactor1977,agarwalNewLeastsquaresRefinement1978},
there for electron densities rather than displacement correlations.

Thermal averaging replaces each atom pair's contribution to this
autocorrelation by a Gaussian centred on its bond vector
$\mathbf{d}_{\kappa\kappa'}^{\vR}=\vR+\vx_\kappa-\vx_{\kappa'}$,
whose covariance is that of the atoms' relative displacement. With a
centred Gaussian written as
\begin{equation}
  \mathcal{N}(\mathbf{r};\boldsymbol{\Sigma})
  =
  \frac{1}{\sqrt{(2\pi)^{3}\det\boldsymbol{\Sigma}}}\,
  \exp\!\Bigl(-\tfrac{1}{2}\,
  \mathbf{r}^{\!\top}\boldsymbol{\Sigma}^{-1}\mathbf{r}\Bigr),
  \label{eq:gaussian_def}
\end{equation}
the correlated pair takes the covariance
$\boldsymbol{\Sigma}_{\kappa\kappa'}^{\vR}$ and the uncorrelated
reference the covariance $\boldsymbol{\Sigma}_{\kappa\kappa'}^{(0)}$,
\begin{equation}
  \begin{aligned}
    \boldsymbol{\Sigma}_{\kappa\kappa'}^{\vR}
    &=\vU_\kappa+\vU_{\kappa'}-2\,\vC_{\kappa\kappa'}(\vR),\\
    \boldsymbol{\Sigma}_{\kappa\kappa'}^{(0)}
    &=\vU_\kappa+\vU_{\kappa'}.
  \end{aligned}
  \label{eq:sigma_def}
\end{equation}
Here $\boldsymbol{\Sigma}_{\kappa\kappa'}^{\vR}$ is the covariance
of the relative displacement
$\vu_{\kappa'}(\vR)-\vu_\kappa(\mathbf{0})$, distinct
from the displacement cross-correlation
$\vC_{\kappa\kappa'}(\vR)=\langle\vu_\kappa(\mathbf{0})
\otimes\vu_{\kappa'}(\vR)\rangle$, and
$\boldsymbol{\Sigma}_{\kappa\kappa'}^{(0)}$ is the same expression with the
correlation $2\vC_{\kappa\kappa'}$ left out. Transforming a
real-space Gaussian of covariance $\boldsymbol{\Sigma}$ with the kernel
$e^{2\pi i\,\vh\cdot\mathbf{r}}$ returns
$\exp(-2\pi^2\,\vh\T\boldsymbol{\Sigma}\vh)$,
recovering the Debye-Waller and correlation factors above.

Each atom carries an electron density $\rho_\kappa(\mathbf{r})$ whose
transform is the form factor, $f_\kappa(\vh)=\mathcal{F}[\rho_\kappa](\vh)$,
so a pair $(\kappa,\kappa')$ contributes the density cross-correlation
$\rho_\kappa\star\rho_{\kappa'}$ smeared by its displacement Gaussian.
The connected autocorrelation, the difference of these two smeared
densities summed over all pairs, is the 3D-$\Delta$PDF
\begin{align}
  \Delta&\mathrm{PDF}(\mathbf{r})
  = \sum_{\vR,\,\kappa,\kappa'}
  \bigl(\rho_\kappa\star\rho_{\kappa'}\bigr)\ast {}
  \nonumber\\
  &\Bigl[
    \mathcal{N}\!\bigl(\mathbf{r}-\mathbf{d}_{\kappa\kappa'}^{\vR};\,
    \boldsymbol{\Sigma}_{\kappa\kappa'}^{\vR}\bigr)
    {}-
    \mathcal{N}\!\bigl(\mathbf{r}-\mathbf{d}_{\kappa\kappa'}^{\vR};\,
    \boldsymbol{\Sigma}_{\kappa\kappa'}^{(0)}\bigr)
  \Bigr],
  \label{eq:delta_pdf_realspace}
\end{align}
where $\ast$ denotes convolution,
so that the all-order intensity of \cref{eq:all_order_diffuse}
reduces to a single Fourier transform~\cite{simonovYellComputerProgram2014},
\begin{equation}
  I_{\mathrm{diff}}(\vh)
  = N\,\mathcal{F}\!\bigl[\Delta\mathrm{PDF}(\mathbf{r})\bigr](\vh),
  \label{eq:idiff_fft_of_blob}
\end{equation}
the convolution supplying the form-factor product
$f_\kappa(\vh) f_{\kappa'}^{*}(\vh)$ on transform.

Each atom pair thus contributes two Gaussians centred on its bond vector,
one from the true pair distribution function, which retains the correlated motion
of the two atoms, and one from the Patterson function of the average structure,
in which the same atoms move independently. Correlation makes the first Gaussian
slightly tighter than the second, so the two do not quite cancel, and this
residue is the diffuse signal.

\subsection{Choice of the sampling grid}

Looking at Eq.~\eqref{eq:delta_pdf_realspace}, one could hope to simply
truncate the pair sum at some radius and keep only the near neighbours.
Unfortunately this does not work. The sharpest features of TDS are the
acoustic peaks at the Bragg positions, and by Fourier duality they
correspond to the slowest-decaying signals in the $\Delta$PDF, namely
the acoustic displacement correlations. Pairs of atoms separated by
tens of unit cells still contribute to the sum, and cutting it off at a
fixed radius casts truncation ripple across the calculated map.

Instead, the set of pairs is fixed by the sampling of reciprocal space.
Sampling the intensity with a step $\Delta q = 1/N$ (say
$1/30$~r.l.u.) defines a supercell of $N^3$ unit cells in PDF space,
and we build the $\Delta$PDF from all pairs within that box. For
consistency, the phonon covariances of
Eq.~\eqref{eq:covariance_from_phonons} are evaluated on the same
$N\times N\times N$ $\vq$-mesh. The inverse DFT then returns
$\vC_{\kappa\kappa'}(\vR)$ periodic on that same box, so the slow
acoustic tails are folded back into the box instead of being cut. The
folding is not free of error, as the correlation of neighbour
$(N{+}2,0,0)$ lands on neighbour $(2,0,0)$, but the misplaced
contributions are the weakest and most distant ones, and the error
decreases as the grid grows. Above all, folding introduces no sharp
edge, so the map stays free of truncation ripple. With one grid serving
both transforms, the $\Delta$PDF and the intensity form an exact DFT
pair and no interpolation is needed.
 \section{Implementation}\label{sec:implementation}

\subsection{Algorithmic structure}

\paragraph{Stage~1: reciprocal-space correlation tensor.}
For phonon branch~$\nu$ at wavevector~$\vq$, frequency
$\omega_{\nu}(\vq)$, and eigenvector component
$e_{\kappa\alpha,\nu}(\vq)$ for atom~$\kappa$ and
Cartesian direction~$\alpha$, the displacement-correlation
tensor is
\begin{equation}
  \gamma_{\kappa\alpha,\kappa'\!\beta}(\vq)
  =
  \frac{1}{\sqrt{M_{\kappa}M_{\kappa'}}}
  \sum_{\nu}
  A_{\nu}(\vq)\,
  \times
  e^{*}_{\kappa\alpha,\nu}(\vq)\,
  e_{\kappa'\!\beta,\nu}(\vq).
  \label{eq:gamma_def}
\end{equation}
This is the reciprocal-space kernel of
Eq.~\eqref{eq:covariance_from_phonons}~\cite{xuXrayThermalDiffuse2010},
up to the constant prefactor $\hbar/2$, which the implementation
applies separately; the $1/N_{\vq}$ normalisation is supplied by the
inverse DFT, which yields $\vC_{\kappa\kappa'}(\vR,T)$ (Stage~2 below).
The thermal amplitude factor
\begin{equation}
  A_{\nu}(\vq)
  =
  \frac{2\langle n_{\nu}\rangle + 1}{\omega_{\nu}(\vq)}
  =
  \frac{1}{\omega_{\nu}(\vq)}
  \coth\!\left(\frac{\hbar\,\omega_{\nu}(\vq)}{2\kB T}\right)
  \label{eq:amp_factor}
\end{equation}
is proportional to the mean-square displacement amplitude of
mode~$\nu$ in the harmonic approximation, with
$\langle n_{\nu}\rangle$ the Bose-Einstein occupation number.

The sum over branches is evaluated simultaneously for all
$\vq$ points via a single \texttt{numpy.einsum}
contraction (Listing~\ref{lst:gamma}).
In the index string \texttt{"hklmsi, hklmtj, hklm, ts -> hklstij"}, 
the letters \texttt{hkl} address the three grid axes, \texttt{m} is 
the branch index~$\nu$, \texttt{s} and~\texttt{t} label the two atoms 
$\kappa,\kappa'$ of the pair, and \texttt{i},~\texttt{j} are the Cartesian 
directions $\alpha,\beta$.
The branch index~\texttt{m} appears in the first three
operands but is absent from the output subscript, so
\texttt{einsum} contracts (sums) over it, directly
implementing the $\sum_\nu$ of Eq.~\eqref{eq:gamma_def}.
The mass-factor matrix \texttt{ts} carries no grid or branch
dependence and is broadcast across all \texttt{hklm} points
at no extra storage cost, equivalent to an outer product with
a $(N_\kappa \times N_\kappa)$ weight matrix applied after the
branch summation.
The output tensor has shape
$(N_1,N_2,N_3,N_\kappa,N_\kappa,3,3)$, with the atom-pair
and Cartesian-pair dimensions kept separate rather than
flattened, so that the result can be directly contracted with
the scattering vector $\vh$ in the subsequent TDS
intensity formula without any index reshaping.

\begin{listing}[!t]
  \caption{Thermal amplitude factor $A_{\nu}(\vq)$
  of Eq.~\eqref{eq:amp_factor}.}
  \label{lst:amp}
\begin{lstlisting}
import numpy as np
NPF64 = np.typing.NDArray[np.float64]

def amp_factor(
    omega: NPF64,         # (N1, N2, N3, n_branches)
    temperature: float = 300.0,
) -> NPF64:
    with np.errstate(divide="ignore"):
        amp = (1.0 / omega) * coth(
            hbar * omega / (2.0 * kb * temperature)
        )
    # acoustic Gamma-point modes: set to zero
    return np.nan_to_num(
        amp, posinf=0, neginf=0, nan=0
    )
\end{lstlisting}
\end{listing}

\begin{listing}[!t]
  \caption{Branch contraction yielding
  $\boldsymbol{\gamma}(\vq)$.}
  \label{lst:gamma}
\begin{lstlisting}
def calc_gamma(
    omega: NPF64,        # (N1, N2, N3, n_branches)
    eigenvectors: NPF64, # (N1, N2, N3, n_branches,
                         #  n_atoms, 3)
    mass_factors: NPF64, # (n_atoms, n_atoms)
                         # entry (t,s) = 1/sqrt(M_t*M_s)
    temperature: float = 300.0,
) -> NPF64:
    amp = amp_factor(omega, temperature)
    # contract over branch index m;
    # retain atom (s,t) and Cartesian (i,j) pairs
    gamma: NPF64 = np.einsum(
        "hklmsi, hklmtj, hklm, ts"
        " -> hklstij",
        np.conj(eigenvectors),
        eigenvectors,
        amp,
        mass_factors,
    )
    return gamma
\end{lstlisting}
\end{listing}

\paragraph{Stage~2: real-space covariance.}
The real-space atomic displacement covariance is obtained by
inverse discrete Fourier transform of $\boldsymbol{\gamma}$ over the three grid axes,
\begin{equation}
  C_{\kappa\alpha,\kappa'\!\beta}(\vR)
  =
  \bigl[\mathcal{F}^{-1}\,\gamma_{\kappa\alpha,\kappa'\!\beta}
  \bigr](\vR).
  \label{eq:covar}
\end{equation}
Because $\boldsymbol{\gamma}(\vq)$ is Hermitian by
construction --- the eigenvectors satisfy
$\ve_{\nu}(-\vq) = \ve^{*}_{\nu}(\vq)$
and $A_\nu$ is real and even in~$\vq$ --- the inverse DFT
over a symmetric grid is real in exact arithmetic.
The \texttt{.real} cast in Listing~\ref{lst:ifft}, implementing
Eq.~\eqref{eq:covar}, discards floating-point imaginary residuals
of order machine epsilon.

\paragraph{Stage~3: diffuse scattering calculation.}
The diffuse intensity is evaluated in \textsc{Yell} by assembling the
3D-$\Delta$PDF and Fourier transforming it
[\cref{eq:delta_pdf_realspace,eq:idiff_fft_of_blob}].
For every atom pair $(\kappa,\kappa')$ and lattice vector $\vR$,
\textsc{Yell} places two centred 3D Gaussians at the interatomic
vector $\mathbf{d}_{\kappa\kappa'}^{\vR}=\vR+\vx_\kappa-\vx_{\kappa'}$,
one with the uncorrelated Patterson covariance
$\boldsymbol{\Sigma}_{\kappa\kappa'}^{(0)}=\vU_\kappa+\vU_{\kappa'}$
and one with the correlated covariance
$\boldsymbol{\Sigma}_{\kappa\kappa'}^{\vR}=
\vU_\kappa+\vU_{\kappa'}-2\,\vC_{\kappa\kappa'}(\vR)$ set by the
Stage-2 covariance.
Their difference is the per-pair
$\Delta$PDF contribution.
Because each Gaussian is sharply localised, the full $\Delta$PDF
volume is cheap to build, and a single FFT of it returns the
all-order diffuse map over the entire grid in one operation,
replacing the per-$\vh$ summation. The form-factor product
$f_\kappa(\vh) f_{\kappa'}^{*}(\vh)$ is applied on transform, reducing
for monatomic silicon to a single $\lvert f(\vh)\rvert^{2}$ envelope on
the diffuse map.

\begin{listing}[!t]
  \caption{Inverse DFT to real-space covariance.}
  \label{lst:ifft}
\begin{lstlisting}
def gamma_to_covar(gamma: NDArray) -> NDArray:
    covar = np.fft.ifftn(gamma, axes=(0, 1, 2))
    covar = covar.real          # discard fp noise
    covar *= 1e20               # m^2 -> Angstrom^2
    return covar
\end{lstlisting}
\end{listing}

\subsection{Numerical details}

\paragraph{Reciprocal-space grid.}
A uniform $N_1\times N_2\times N_3$ grid spanning one
reciprocal unit cell is used.
Acoustic modes at $\vq=\boldsymbol{0}$ are regularised to
zero, consistent with the zero net-force condition for
rigid-body translation (see Supplementary Information).

\paragraph{\textit{Ab initio} phonon model.}
Harmonic interatomic force constants (IFCs) for silicon were computed
from density-functional perturbation theory (DFPT) using
\textsc{Abinit}~\cite{gonzeAbinitprojectImpactEnvironment2020}.
The exchange-correlation functional was GGA-PBE;
the ionic cores were described by a norm-conserving pseudopotential
(\texttt{Si.psp8}, ONCVPSP format, ABINIT pseudopotential table).
The cell volume was optimised by variable-cell BFGS relaxation
(\texttt{optcell}=1, volume only), yielding
$a = \SI{5.469}{\angstrom}$ (experimental: $\SI{5.431}{\angstrom}$,
$+0.7\%$, consistent with the GGA overestimation trend).
The plane-wave cutoff was $E_{\mathrm{cut}} = \SI{20}{\hartree}$
(\SI{544}{\electronvolt}) and the Brillouin zone was sampled on a
$\Gamma$-centred $14\times14\times14$ Monkhorst-Pack $\vk$-mesh;
both were validated by total-energy convergence tests
(Figure~S1), with errors $<\SI{0.05}{\meV\per atom}$ and
$<\SI{0.02}{\meV\per atom}$, respectively.
Dynamical matrices were computed by DFPT on a $14\times14\times14$
$\vq$-mesh.
The acoustic sum rule was enforced (\texttt{asr}=2) and long-range
quadrupole-quadrupole interactions were included
(\texttt{dipqua}=1, \texttt{quadqu}=1) while the dipole-dipole
term was omitted (\texttt{dipdip}=0), appropriate for silicon
which has no LO-TO splitting.
Phonon frequencies and eigenvectors were interpolated by
\textsc{anaddb}~\cite{gonzeAbinitprojectImpactEnvironment2020}
onto a $60\times60\times60$ $\vq$-grid for subsequent
covariance-volume assembly.
The intensity step of $\tfrac{1}{30}$~r.l.u.\ used below takes every
second point of this grid; such an integer subsampling keeps the two
transforms commensurate and requires no interpolation.

\paragraph{Performance and memory.}
The cost is dominated by the single Fourier transform of the
$\Delta$PDF over the target reciprocal-space grid. For the silicon
benchmark on a $601^3$ grid ($\pm 10$ r.l.u., step
$\tfrac{1}{30}$ r.l.u., $\sim$\num{2.2e8} voxels), assembling the
real-space covariance volume $\vC(\vR)$ takes
\SI{48}{\second} and the full diffuse map is then obtained in a
single \SI{532}{\second} FFT evaluation, under \SI{10}{\minute}
end to end on a single workstation (Supplementary Information).
Because $\vC(\vR)$ is stored only on the small
real-space supercell that supports the displacement correlations,
its memory footprint is negligible compared with the full
reciprocal-space output volume, which dominates storage.

\subsection{Software and reproducibility}

The implementation is written in Python~3.10+ and depends on
NumPy~\cite{harrisArrayProgrammingNumPy2020} and SciPy~\cite{virtanenSciPy10Fundamental2020}.
Array shapes and scalar types are annotated via
\texttt{numpy.typing} aliases (\texttt{NPF64},
\texttt{NPC128}, etc.) to make operand rank explicit at each
call site.
 \begin{figure}[!t]
  \centering
  \includegraphics[width=\figw]{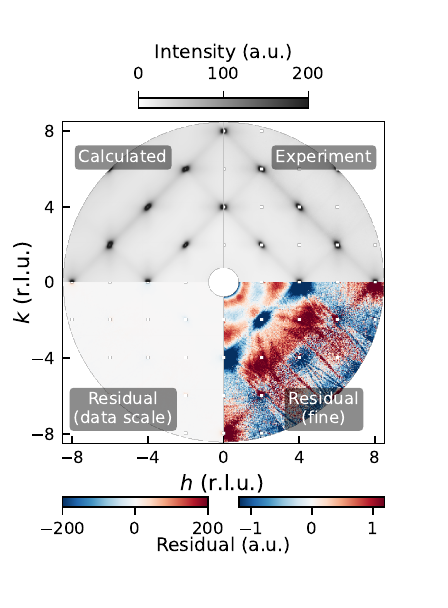}
  \caption{\textbf{HK0 section of the silicon TDS volume at
    $T = \SI{293.15}{\kelvin}$, Bragg-punch radius $r_p = 3$}
    (ID28, ESRF).
    Top left, calculated intensity
    $s\,I^{\mathrm{calc}}$ plus refined background; top right, measured
    intensity (greyscale). Bottom left, residual on the data scale
    ($\pm200$ a.u.); bottom right, the same residual on a fine scale
    ($\pm1$ a.u.), with a diverging blue/white/red map for
  negative/zero/positive values. Bragg peaks are masked.}\label{fig:hk0_quadrant}
\end{figure}

\section{Validation}\label{sec:validation}

We selected silicon as the benchmark for this study, as it is a cubic
crystal with only two atoms in the primitive unit cell, no disorder,
and a well-characterised phonon dispersion.

Single-crystal diffraction data were collected at the ESRF ID28
beamline using monochromatic X-rays of wavelength
\SI{0.6968}{\angstrom} (\SI{17.79}{\kilo\electronvolt}), focused to a
50\,\textmu{}m beam. The sample, a square prism of roughly
200\,\textmu{}m in cross-section and a few millimetres long, was glued
to a glass capillary with epoxy and mounted on the goniometer. Data
were recorded at room temperature ($T = \SI{293.15}{\kelvin}$) by
rotating the crystal $360^{\circ}$ about the $\phi$ axis in
$0.1^{\circ}$ steps (3600 frames). To broaden the reciprocal-space
coverage of the Pilatus 1M detector, two datasets were taken with the
detector at $19^{\circ}$ and $48^{\circ}$. Raw frames were indexed
with XDS~\cite{kabschXDS2010} and each dataset reconstructed into a
three-dimensional reciprocal-space volume with
\textsc{Meerkat}~\cite{simonovMeerkat} on a $\tfrac{1}{30}$~r.l.u.\
grid, correcting for the absorption efficiency of the silicon sensor
layer in the Pilatus detector. The two volumes were then merged and
symmetry-averaged over the $m\bar{3}m$ Laue group to improve the
statistics of the diffuse scattering signal.

As a second, independent test we measured a temperature series of
silicon at the Swiss-Norwegian Beamlines (SNBL/BM01, ESRF;
$\lambda = \SI{0.71943}{\angstrom}$),
collecting data at $200$, $250$, $300$ and \SI{350}{\kelvin} and
reconstructing each dataset in the same way on a finer
$\tfrac{1}{60}$~r.l.u.\ grid. The silicon crystal was measured as a
side sample during a study of BaTi$X_3$, and its data are archived with
that session~\cite{bulledBaTiX2026}.

Both datasets were analysed identically. The TDS intensity was
calculated from the \textit{ab initio} phonon model at the sample
temperature, with the force constants used without adjustment; only an
overall scale factor and a smooth polynomial background were refined,
by solving a linear least-squares problem. Bragg peaks were removed by
punch-and-fill with radius $r_p$ before comparison.

\begin{figure}[!t]
  \centering
  \includegraphics[width=\figw]{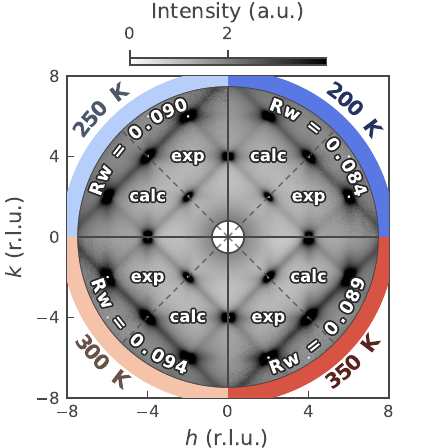}
  \caption{\textbf{Silicon $HK0$ thermal diffuse scattering at four
    temperatures} (SNBL/BM01, ESRF).
    Each $45^\circ$ sector shows the measured (\textit{exp}) or calculated
    (\textit{calc}, $s\,I^{\mathrm{calc}}$ plus refined background) intensity for a
    single temperature; the outer ring encodes temperature
    ($200$--\SI{350}{\kelvin}, blue to red) and the per-temperature
    residual $R_2$. Bragg peaks are removed by punch-and-fill ($r_p=4$); the
  greyscale is linear in intensity.}\label{fig:snbl_composite}
\end{figure}

For the ID28 benchmark the model reproduces the measured diffuse
scattering to $R_2 \approx 4.8\%$ ($r_p = 3$; \cref{fig:hk0_quadrant}).
The fitting procedure and the punch-radius grid search are given in the
Supplementary Information.
For the temperature series the model reproduces the diffuse scattering
at every temperature with $R_2 \approx 9\%$
($r_p = 4$; \cref{fig:snbl_composite}). A single, temperature-independent scale
factor was used for the whole series, so the thermal growth of the
intensity is carried by the model itself rather than by a change of
scale. The higher residual relative to ID28 reflects the lower
signal-to-noise of the diffuse scattering and a stricter Bragg punch,
as the finer grid makes a given punch radius physically smaller. The
temperature-difference maps and fit summary are given in the
Supplementary Information.

 \section{Discussion}

With no structural or force-constant parameters refined, the model
reproduces the measured diffuse scattering to $R_2 \approx 4.8\%$.
For diffuse scattering, a signal orders of magnitude weaker than the
Bragg peaks it accompanies, this is a remarkably low residual, and it
measures the combined accuracy of the harmonic phonon model and the
all-order intensity formula rather than the quality of a fit.

The origin of the remaining $4.8\%$ is difficult to pin down, and
the largest part of it likely lies in the data processing rather than in the
model. On the model side, several effects plausibly contribute at
this level. The calculation is purely harmonic, and for silicon at
room temperature the harmonic approximation is accurate but not
exact, so some anharmonic scattering is left unreproduced. The force
constants are computed within GGA-PBE, whose $0.7\%$ overestimate of
the lattice constant (\cref{sec:implementation}) softens the phonon
frequencies by roughly $2\%$ through the Gr\"uneisen effect. The
background is modelled as a single smooth, isotropic function of
$|\vh|$, which absorbs the nearly isotropic Compton and
sample-environment scattering but not any genuinely anisotropic
component. Finally, the largest residuals sit near the Bragg peaks,
where the intensity is plausibly still dominated by the linear
one-phonon term and the folded acoustic correlations are least
accurate; but this is also where the signal is strongest and the
errors of both data and model are therefore largest, so the
attribution remains uncertain. It is this contribution that is
removed by increasing $r_p$.

The closest existing approach is the special displacement method of
Zacharias
\textit{et al.}~\cite{zachariasEfficientFirstPrinciplesMethodology2021,
zacharias2021multiphonon}, which reaches the same all-order harmonic
physics through the diffuse scattering of a single, carefully
displaced supercell. The present scheme has three practical
advantages. The covariance map is assembled directly and cheaply from
the phonon eigenvectors, and the whole diffuse map follows in a
single FFT. Every step of the construction is differentiable, which
opens the way to gradient-based refinement. And the Laue symmetry of
the crystal reduces the number of pairs to be built, by a factor of
48 for silicon. The price is the number of signals. The $\Delta$PDF
assembly handles of order $M^2 N_1 N_2 N_3$ pair signals (reduced by
symmetry) for $M$ atoms in the unit cell, whereas a
special-displacement supercell needs only $M N_1 N_2 N_3$ displaced
atoms; we doubt, however, that diffuse-scattering studies will soon
reach system sizes where this difference decides. Beyond cost, the
present scheme outputs a \ddPDF{} that can be refined jointly with
static disorder in the same framework, rather than a result
disconnected from the real-space picture.
 \section{Conclusions}

We have presented a method that calculates X-ray thermal diffuse
scattering to all phonon orders, by building the real-space
displacement covariance from first-principles phonons and Fourier
transforming the resulting 3D-$\Delta$PDF. A single FFT delivers the
full diffuse map, which makes the calculation practical for the large
reciprocal-space volumes collected at modern diffractometers.

We validated the method on silicon. With nothing refined beyond a
scale and a background, the calculation matches the measured diffuse
scattering to $R_2 \approx 4.8\%$.

Finally, because the method uses the same pair-correlation language
as the \textsc{Yell} refinement of static disorder, thermal and
static correlations can now be refined jointly, removing a principal
obstacle in the analysis of crystals where the two coexist.

 \begin{acknowledgements}
The authors thank the European Synchrotron Radiation Facility for
provision of beamtime at the ID28 and BM01 (SNBL) beamlines, and
Alexei Bossak for assistance with the ID28 experiment and discussions.
During the preparation of this manuscript the authors used Claude
(Anthropic) to assist with language editing. The authors reviewed and
edited all suggestions and take full responsibility for the content of
the publication.
\end{acknowledgements}

\begin{funding}
This work was supported by the Swiss National Science Foundation (SNSF)
under grant No.\ 203658.
\end{funding}

\ConflictsOfInterest{The authors declare no conflicts of interest.}

\DataAvailability{The ID28 dataset, together with the source code, \textit{ab initio}
inputs, and a reproduction script, is archived in the ETH Library
repository and after short review will be available on \href{http://hdl.handle.net/20.500.11850/803367}{\nolinkurl{http://hdl.handle.net/20.500.11850/803367}} as well as registered on doi \href{https://doi.org/10.3929/ethz-c-000803367}{\nolinkurl{10.3929/ethz-c-000803367}}.
The SNBL/BM01 silicon temperature-series data are archived in the ESRF
Data Repository under
\href{https://doi.org/10.15151/ESRF-ES-2312167790}{\nolinkurl{10.15151/ESRF-ES-2312167790}}~\cite{bulledBaTiX2026}.
}
\bibliography{references}

\end{document}